\documentclass[%
 reprint,
 superscriptaddress,
showpacs,preprintnumbers,
 amsmath,amssymb,
 aps,
 prb,
 longbibliography,
]{revtex4-1}

\usepackage{graphicx}
\usepackage{dcolumn}
\usepackage{bm}
\usepackage{color}
\usepackage{ulem}

\begin{document}

\preprint{APS/123-QED}

\title{
A strong coupling theory for electron-mediated interactions\\
in double-exchange models
}

\author{Hiroaki Ishizuka}
\affiliation{
Kavli Institute for Theoretical Physics, University of California, Santa Barbara, California 93106, USA
}

\author{Yukitoshi Motome}
\affiliation{
Department of Applied Physics, University of Tokyo, Hongo, 7-3-1, Bunkyo, Tokyo 113-8656, Japan
}

\date{\today}

\begin{abstract}
We present a theoretical framework for evaluating effective interactions between localized spins mediated by itinerant electrons in double-exchange models. Performing the expansion with respect to the spin-dependent part of the electron hopping terms, we show a systematic way of constructing the effective spin model in the large Hund's coupling limit. As a benchmark, we examine the accuracy of this method by comparing the results with the numerical solutions for the spin-ice type model on a pyrochlore lattice. We also discuss an extension of the method to the double-exchange models with Heisenberg and $XY$ localized spins.
\end{abstract}

\maketitle

\section{Introduction}

Spin-charge coupled systems, which consist of itinerant electrons interacting with localized moments, are of special interest in the study of itinerant magnets for their rich physics from the interplay between magnetic moments and itinerant electrons. A key aspect that gives rise to the rich physics is the effective interactions between localized moments mediated by the itinerant electrons. In the weak coupling limit, the itinerant electrons induce long-ranged effective exchange interactions with oscillating signs.~\cite{Ruderman1954,Kasuya1956,Yosida1957} The weak coupling theory of effective spin interactions, so-called Ruderman-Kittel-Kasuya-Yosida (RKKY) interactions, has achieved a success in the study of metallic magnets, such as transition-metal~\cite{Kasuya1956} and rare-earth~\cite{Yosida1957} compounds and spin-glass behavior in alloys lightly doped with magnetic ions.~\cite{Binder1986}

On the other hand, a strong ferromagnetic coupling between localized moments and itinerant electrons may arise in the transition metal systems due to the strong Hund's coupling. In the strong coupling limit, the spins of itinerant electrons are fully polarized along the localized moments. Such a situation is well described by the double-exchange (DE) model. In this model, the kinetic motion of electrons induces an effective ferromagnetic (FM) interaction between localized moments.~\cite{Zener1951,Anderson1955} This is called the DE interaction, which stabilizes a metallic FM state at low temperature. In these oxides, however, there generally exists antiferromagnetic (AFM) super-exchange (SE) interaction between the localized moments as well, and the competition of two interactions may give rise to nontrivial phenomena. One such example is found in perovskite manganese oxides. They are renown for the colossal magnetoresistance,~\cite{Tokura2000,Salamon2001,Dagotto2001} in which inhomogeneity in the competing region between FM and AFM phases have been studied in relation to the magnetoresistance.~\cite{Uehara1999,Moritomo1999,Moreo1999,Balagurov2001,Lai2010}

While the simple picture based on the above argument appears to work well in the manganese oxides, recent numerical studies of DE models on geometrically frustrated lattices~\cite{Kumar2010,Venderbos2012,Ishizuka2013,Reja2015} have discovered emergence of intermediate phases in the competing region. In the case of checkerboard and triangular lattices, instabilities toward noncoplanar spin orderings were observed in the Monte Carlo (MC) simulation for the models with localized classical Heisenberg moments.~\cite{Kumar2010,Venderbos2012} Meanwhile, a thermally-induced intermediate phase with spontaneously-broken spatial inversion symmetry was found in the model with Ising moments on a pyrochlore lattice.~\cite{Ishizuka2013} These results imply that, in the frustrated systems, the subdominant interactions beyond the simple DE mechanism potentially give rise to the nontrivial phases in the phase competing region. Indeed, in the previous study by the authors,~\cite{Ishizuka2013} effective further-neighbor interactions derived from a strong coupling theory successfully predicted the presence of the intermediate phase. However, the method of the strong coupling expansion was not described in detail, and its accuracy has not been examined systematically.

\begin{figure}
   \includegraphics[width=\linewidth]{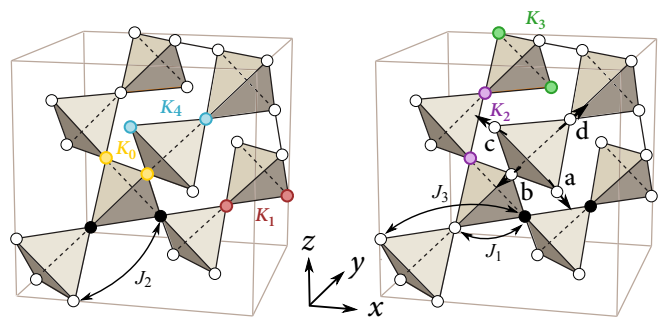}
   \caption{(Color online).
   Schematic picture of a pyrochlore lattice. Each figure indicates the two-spin exchange interactions, $J_l$, and the four-spin interactions, $K_l$ ($l$ is an integer). For the latter, four Ising spins at the two colored sites $\langle k, l \rangle$ and two black sites $\langle i,j \rangle$ interact with each other. The arrows on the sites in the right panel indicate anisotropy axes of the Ising moments, $\bm{n}_i$. See the text for details.
   }
   \label{fig:lattice}
\end{figure}

In this paper, we present the framework of the strong coupling theory for the effective spin interactions between the localized moments in DE models. We illustrate the technique for a spin-ice type DE model on a pyrochlore lattice (Fig.~\ref{fig:lattice}). We show that the second-order expansion gives rise to various four-spin interactions in addition to the two-spin interactions between second- and third-neighbor sites. The results for the effective spin model are compared with numerical results for the DE model, to test the accuracy of this theory. We find that the expansion up to second order correctly reproduces the trend of magnetic phases while changing the AFM SE interaction and the itinerant electron density. 

The organization of this paper is as follows. In Sec.~\ref{sec:m&m}, we introduce the model we consider, and elaborate the details of strong coupling expansion used here. In Sec.~\ref{sec:result}, we investigate the accuracy of this method by comparing the results with those of numerical diagonalization and MC simulation. In the last, an extension of our theory to Heisenberg and $XY$ localized moments is briefly discussed in Sec.~\ref{sec:heisenberg}. Section~\ref{sec:summary} is devoted to discussions and summary of this paper.

\section{Model and Method}\label{sec:m&m}

In this section, we introduce the model and method we used. In Sec.~\ref{ssec:m&m:model}, we introduce the DE model we consider in this paper. We explain the strong coupling expansion method in Sec.~\ref{ssec:m&m:method}.

\subsection{Model}\label{ssec:m&m:model}

The DE model we consider in this paper consists of itinerant electrons and classical localized moments that are strongly coupled to each other. The Hamiltonian is given in the general form~\cite{Zener1951,Anderson1955} 
\begin{eqnarray}
H = - \sum_{i,j} \! t_{ij} c^\dagger_{i} c_{j} + \frac12\sum_{i,j}J_{ij}{\bm S}_i\cdot{\bm S}_j, \label{eq:HDE}
\end{eqnarray}
where $c_i$ ($c^\dagger_i$) is the annihilation (creation) operator of an itinerant electron and ${\bm S}_i$ is the localized moment at $i$th site. The electrons are described by spinless fermions, as their spins are perfectly polarized parallel to the localized spins. The first sum is the kinetic term of itinerant electrons. The transfer integral $t_{ij}$ depends on the relative position of $i$th and $j$th sites, and also depends on the localized spins at the two sites:~\cite{mHartmann1996}
\begin{eqnarray}
t_{ij}&=& t^\text{(bare)}_{ij} \left\{ \cos\frac{\theta_i}2\cos\frac{\theta_j}2\right.\nonumber\\
      & &\left.\quad\qquad+ \sin\frac{\theta_i}2\sin\frac{\theta_j}2\exp[\mathrm{i}(\phi_j-\phi_i)] \right\}.\label{eq:transfer}
\end{eqnarray}
Here, $\theta_i$ and $\phi_i$ are the polar and azimuthal angles of ${\bm S}_i$, respectively; $t^\text{(bare)}_{ij}$ is the transfer integral between $i$th and $j$th sites in the absence of the coupling to localized spins. The second sum in Eq.~(\ref{eq:HDE}) is the AFM SE interaction term in which $J_{ij}$ is the exchange coupling between $i$th and $j$th sites. In this paper, we particularly consider the case of Ising local moments, i.e., ${\bm S}_i=\pm {\bm n}_i$ with ${\bm n}_i$ being a unit vector parallel to the anisotropy axis at $i$th site.

\subsection{Strong coupling expansion}\label{ssec:m&m:method}

To investigate the magnetic properties of the model in Eq.~(\ref{eq:HDE}), understanding the nature of effective spin interactions mediated by the coupled fermions is of crucial importance. To evaluate the effective spin interactions, we start by approximating the transfer integral in Eq.~(\ref{eq:transfer}) by its amplitude,
\begin{eqnarray}
\frac{\tilde{t}_{ij}}{|t^\text{(bare)}_{ij}|}= \left|\frac{t_{ij}}{t^\text{(bare)}_{ij}}\right| = \sqrt{\frac{1+\cos\theta_{ij}}2}\label{eq:transfer2},
\end{eqnarray}
where $\theta_{ij}$ is the angle between $\bm{S}_i$ and $\bm{S}_j$.~\cite{note_nophase} The DE model with this approximated form of the transfer integral, $\tilde{t}_{ij}$, was used to study the magnetic properties of manganese oxides in several previous works.~\cite{deGennes1959,Kubo1972,Millis1995,Calderon1998} Using this approximation, and considering that the localized moments are of Ising type, we can rewrite the transfer integral as
\begin{eqnarray}
\tilde{t}_{ij}= t_{ij}^0 + t_{ij}^1 \tilde{S}_i \tilde{S}_j
\label{eq:tilde_tij}
\end{eqnarray}
where $\tilde{S}_i=\bm{S}_i \cdot \bm{n}_i =\pm1$ is the projected spin parameter along ${\bm n}_i$, and
\begin{eqnarray}
t_{ij}^{0} &=& \frac{|t_{ij}^\text{(bare)}|}2 \left( \cos\frac{\theta^0_{ij}}2 + \sin\frac{\theta^0_{ij}}2 \right),\\
t_{ij}^{1} &=& \frac{|t_{ij}^\text{(bare)}|}2 \left( \cos\frac{\theta^0_{ij}}2 - \sin\frac{\theta^0_{ij}}2 \right),
\label{eq:t1ij}
\end{eqnarray}
are real coefficients. Here, $\theta^0_{ij}$ is the relative angle between ${\bm n}_i$ and ${\bm n}_j$.

We consider the hopping term with the coefficient $t_{ij}^0$ as the unperturbed Hamiltonian, $H_0$, and perform the expansion of Matsubara Green's function with respect to the remaining term with $t_{ij}^1$, $H_1$. The Dyson equation is given by
\begin{align}
G_{i,j}({\rm i}\omega)=g_{i,j}({\rm i}\omega)-\sum_{k,l} g_{i,k}({\rm i}\omega) \left[\,t_{kl}^1\tilde{S}_{k} \tilde{S}_{l}\,\right] G_{l,j}({\rm i}\omega).\label{eq:Dyson}
\end{align}
Here, $G_{i,j}({\rm i}\omega)$ is Matsubara Green's function and $g_{i,j}({\rm i}\omega)$ is bare Green's function of the unperturbed Hamiltonian. The term in the square bracket in Eq.~(\ref{eq:Dyson}) is the scattering by $H_1$.
The internal energy of the system is given by
\begin{eqnarray}
E = - \sum_{i,j} \! \tilde{t}_{ij} \langle c^\dagger_{i}c_{j}\rangle+\frac12\sum_{i,j} \! J_{ij} {\bm S}_{i}\cdot{\bm S}_{j}. \label{eq:Eint}
\end{eqnarray}
By replacing $\langle c^\dagger_{i}c_{j}\rangle$ by $\sum_\omega G_{j,i}({\rm i}\omega)e^{{\rm i}\omega(-0)}$ and expanding Green's functions using the Dyson equation, one obtains the effective spin model: the energy in Eq.~(\ref{eq:Eint}) gives the effective Hamiltonian for the Ising spins $\tilde{S}_i$. In this paper, we consider the expansion up to $O[(t_{ij}^1)^2]$, which leads to effective four-spin interactions in addition to two-spin ones.

Regarding to the accuracy of this method, we note that in Eq.~(\ref{eq:t1ij}), $|t_{ij}^1|$ becomes small when $\theta^0_{ij}$ is close to $\pi/2$. Hence, it is expected that the perturbation is expected to be accurate when $\theta^0_{ij} \sim \pi/2$, namely, the local anisotropy axes are perpendicular to each other. On the other hand, the approximation becomes less accurate as we approach the collinear case, $t_{ij}^{0}=t_{ij}^{1}$.

In the following sections, we test this method for a DE model on a pyrochlore lattice with only nearest-neighbor (NN) transfer integrals and the localized Ising moments having spin-ice type anisotropy.~\cite{Harris1997,Ramirez1999} In this model, the anisotropy axes of two NN spins have the relative angle of $\theta_{ij}^0\sim109^\circ$, which is close to $\pi/2$.

\section{Results}\label{sec:result}

In this section, as the benchmark of the method in the previous section, we study the effective spin interactions in a spin-ice type DE model on a pyrochlore lattice. In Sec.~\ref{ssec:result:sct}, we present the effective spin model obtained from the strong coupling expansion.
The accuracy of this method is investigated by comparing the ground state energy (Sec.~\ref{ssec:result:compete}) and magnetic phase diagram for $n$ and $J$ (Sec.~\ref{ssec:result:variational}). The later is obtained by a variational method. The relevance of the phase diagram is further investigated in Sec.~\ref{ssec:result:mc} by a MC method.

\subsection{Effective spin interactions} \label{ssec:result:sct}

\begin{figure}
   \includegraphics[width=0.92\linewidth]{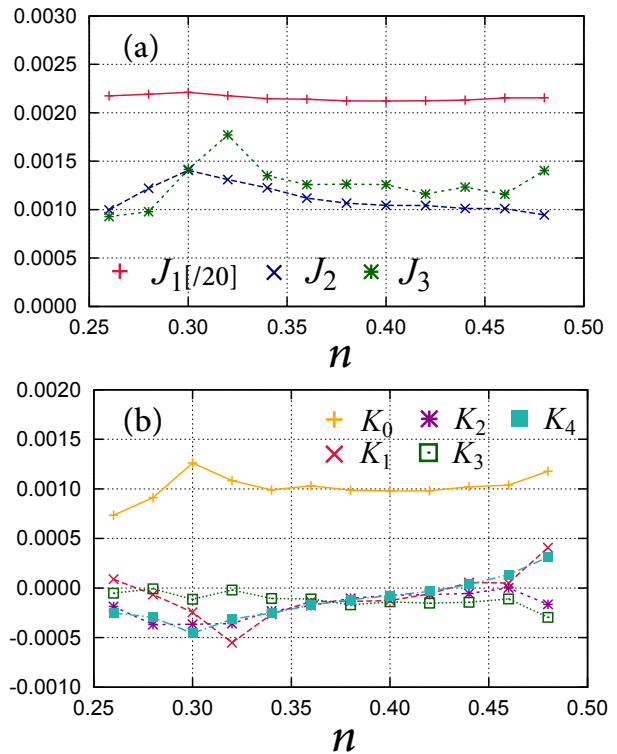}
   \caption{(Color online).
   \label{fig:JsKs}
   Effective spin interactions mediated by itinerant electrons: (a) two-spin and (b) four-spin interactions as functions of the electron density $n$. The definition of interactions is given in Fig.~\ref{fig:lattice}.
   }
   \label{fig:interaction}
\end{figure}

To test the accuracy of the strong coupling theory, we consider an Ising spin DE model on a pyrochlore lattice with hoppings and SE interactions for NN sites only. The anisotropy axes for Ising moments are defined along local $[111]$ axes: ${\bm n}_\text{a}=(1/\sqrt3)(1,-1,-1)$, ${\bm n}_\text{b}=(1/\sqrt3)(-1,1,-1)$, ${\bm n}_\text{c}=(1/\sqrt3)(-1,-1,1)$, and ${\bm n}_\text{d}=(1/\sqrt3)(1,1,1)$ (see Fig.~\ref{fig:lattice}). For simplicity, we replace the NN hopping integral $t_{ij}$ by $\tilde{t}_{ij}$ as in Eq.~(\ref{eq:tilde_tij}); the starting Hamiltonian is given by
\begin{eqnarray}
H = - \sum_{\langle i,j\rangle} \! \tilde{t}_{ij}(c^\dagger_{i} c_{j} + \text{H.c.})+J\sum_{\langle i,j\rangle}{\bm S}_i\cdot{\bm S}_j, \label{eq:HDE2}
\end{eqnarray}
where the sum $\langle i,j\rangle$ is taken over the NN sites on the pyrochlore lattice. Hereafter, we take the bare hopping for NN sites, $t=1$, as the energy unit. 

Applying the strong coupling expansion in Sec.~\ref{ssec:m&m:method} to this model, we construct an effective Ising model on the pyrochlore lattice, whose Hamiltonian is given by
\begin{align}
		  H_\text{eff} &= \sum_{i,j} \! J_{ij} \tilde{S}_i \tilde{S}_j + \frac12\sum_{\substack{\langle i,j\rangle,\langle k,l\rangle,\\ i, j\ne k, l}}\!K_{\langle i,j\rangle,\langle k,l\rangle}\tilde{S}_i\tilde{S}_j\tilde{S}_k\tilde{S}_l\nonumber \\
&-\frac{J}3\sum_{\langle i,j\rangle} \! \tilde{S}_i \tilde{S}_j,\label{eq:Heff}
\end{align}
up to a constant given by the contribution from $H_0$. Here, the first term represents effective two-spin interactions for nearest-, second- and third-neighbor sites, while the second term describes the four-spin interactions. The last term is the AFM SE interaction already present in Eq.~(\ref{eq:HDE2}), where the coefficient $-1/3$ comes from the projection of the Ising spins to the anisotropy axes at each site. In the model in Eq.~(\ref{eq:HDE2}), the leading order in expansion gives NN two-spin interaction while the second order gives four spin interactions between spins on the edge of two bonds at arbitrary distance. The second and third neighbor two-spin interactions also arise from the second order expansion, when the two bonds share a site, e.g., $j=k$. In general, the two-spin interactions between sites with $l$ Manhattan distance arise from the $l$th order in expansion, i.e., they decay exponentially with distance in contrast to the RKKY interaction.

Figure~\ref{fig:interaction} shows the effective spin interactions that arise from the expansion for the model in Eq.~(\ref{eq:HDE2}) as functions of the itinerant electron density $n=\sum_i\langle c_i^\dagger c_i \rangle/N$ ($N$ is the number of sites). We here show the results for $1/4 < n < 1/2$, as the higher order terms appear to be relevant for lower $n$ (see below) and the flat band may lead to some complexity for higher $n$. The definition of each interaction is shown in Fig.~\ref{fig:lattice}. As shown in Fig.~\ref{fig:interaction}, the most dominant interaction mediated by itinerant fermions is the FM NN interaction $J_1$, consistent with what is expected in DE models. This dominant interaction gives rise to instability toward a FM state when the SE interaction $J$ between localized moments is sufficiently weak.

In addition, we also have other two-spin and four-spin interactions which are an order of magnitude smaller than $J_1$. These interactions potentially become important in the phase competing region between DE-driven FM and SE driven AFM phases, i.e., when the SE interaction cancels the NN FM interaction. Some of these subdominant interactions are also plotted in Fig.~\ref{fig:interaction}. In this density range, most of the four-spin interactions decay rapidly with distance; $K_0$ is the interaction of four spins on the same tetrahedron, while $K_i$ ($i\ne0$) are interactions between spins with further distance. As a consequence, the dominant interactions are $J_2$, $J_3$, and $K_0$ for electron densities $n\gtrsim1/4$. The result implies that, for $n\gtrsim1/4$, only considering a limited number of interactions are sufficient in reproducing the qualitative nature of the model.

\subsection{Numerical diagonalization} \label{ssec:result:compete}

\begin{figure}
   \includegraphics[width=\linewidth]{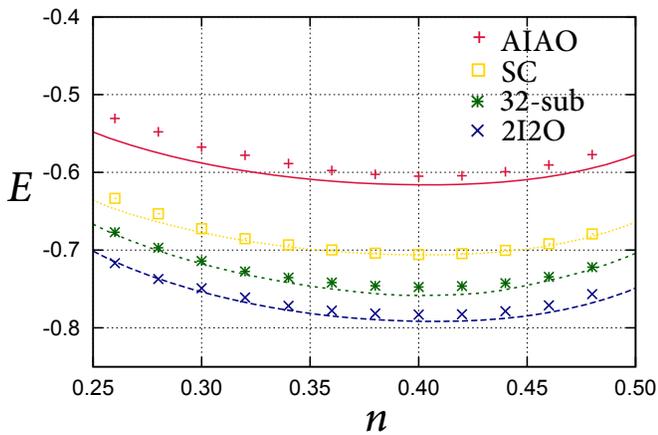}
   \caption{(Color online).
   $n$ dependence of the ground state energy for different spin configurations at $J=0$. The lines are the results obtained by numerical diagonalization of the DE model in Eq.~(\ref{eq:HDE2}), and the symbols are those for the effective spin model in Eq.~(\ref{eq:Heff}) obtained from the strong coupling expansion. See the text for details.
   }
   \label{fig:Egs}
\end{figure}

To examine the accuracy of above theory, we start by evaluating the ground state energy. The results are compared to those of numerical diagonalization for several different spin configurations. We here consider $q=0$ orders with all spins on a tetrahedron pointing inward or outward (all-in/all-out; AIAO) and two spins inward and two spins outward (two-in two-out; 2I2O) as an example of the magnetic order with AFM and FM NN correlation, respectively. In addition to these phases, two magnetic orders that were found in the related Kondo lattice models, 32-sublattice (32-sub) order~\cite{Ishizuka2012} and spin cluster (SC) order,~\cite{Ishizuka2013} are also considered.

Figure~\ref{fig:Egs} shows the result of the ground state energy calculated for different magnetic orders in the effective spin model in Eq.~(\ref{eq:Heff}) by taking into account $J_1$, $J_2$, $J_3$, and $K_0$. For $n\gtrsim1/4$, the results of strong coupling expansion are in accordance with those of numerical diagonalization for the model in Eq.~(\ref{eq:HDE2}). As the NN FM interaction $J_1$ gives the largest contribution to the ground state energy, this result shows that the estimate of $J_1$ is successful in the strong coupling expansion.

We also calculated the ground state energy for $n\lesssim1/4$. In this region, however, the results show strong deviation from the result of numerical diagonalization (not shown here). This is presumably due to the presence of longer-range interactions than second- and third-neighbors which are more dominant than the case for $n\gtrsim1/4$.

\subsection{Variational phase diagram} \label{ssec:result:variational}

\begin{figure}
   \includegraphics[width=\linewidth]{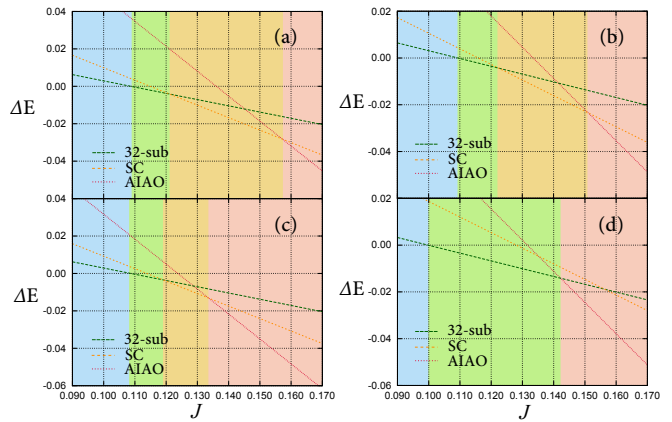}
   \caption{(Color online).
   $J$ dependence of the ground state energy for different spin configurations calculated for the effective spin model in Eq.~(\ref{eq:Heff}) obtained from the strong coupling theory [(a) and (b)] and the DE model in Eq.~(\ref{eq:HDE2}) by numerical diagonalization [(c) and (d)]. The energies are measured from those for the 2I2O state. The electron density were set at $n=0.3$ ($0.4$) for (a) and (c) [(b) and (d)].
   }
   \label{fig:Egs2}
\end{figure}

\begin{figure*}
   \includegraphics[width=\linewidth]{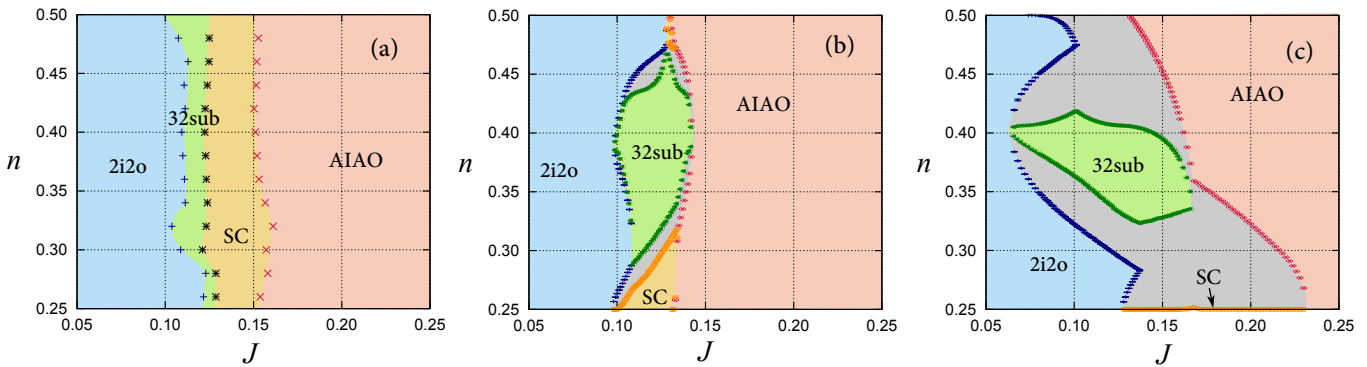}
   \caption{(Color online).
   Variational phase diagrams for (a) the effective spin model obtained by the strong coupling expansion [Eq.~(\ref{eq:Heff})], (b) the DE model in Eq.~(\ref{eq:HDE2}), and (c) the DE model in Eq.~(\ref{eq:HDE}). The phase diagram in (a) is calculated by only considering $J_1$, $J_2$, $J_3$, and $K_0$.
   }
   \label{fig:Jdiag}
\end{figure*}

Next, to evaluate the effect from further-neighbor two-spin interactions and four-spin interactions in Eq.~(\ref{eq:Heff}), we study the ground state phase diagram by a variational calculation with changing the AFM SE interaction $J$. As $J$ tends to cancel the effective NN FM interaction from the DE mechanism, this essentially corresponds to studying the effect of subdominant interactions that arise from the second order expansion in the strong coupling theory. For this purpose, we here perform a variational calculation comparing the ground state energy for different spin configurations: 2I2O, 32-sub, SC, and AIAO orders.~\cite{note_variational}

Figure~\ref{fig:Egs2} shows the results of ground state energy for different magnetic orders measured from that for the 2I2O state. Figure~\ref{fig:Egs2}(a) shows the result for the effective spin model in Eq.~(\ref{eq:Heff}) obtained by the strong coupling theory at $n=0.3$. We find all four phases with increasing $J$; the transition takes place from the 2I2O to 32-sub state at $J\sim0.109$, to SC state at $J\sim0.121$, and to AIAO state at $J\sim0.157$. A similar trend is also found in the result calculated by numerical diagonalization of the model in Eq.~(\ref{eq:HDE2}) [Fig.~\ref{fig:Egs2}(c)]. Note that the electronic phase separation is not considered here; see Fig.~\ref{fig:Jdiag}.

In the results for $n=0.4$, the strong coupling theory predicts four magnetic phases in the ground state, as shown in Fig.~\ref{fig:Egs2}(b). On the other hand, as shown in Fig.~\ref{fig:Egs2}(d), there are only three phases for the model in Eq.~(\ref{eq:HDE2}): 2I2O, 32-sub, and AIAO phases.

Performing the above variational calculation while varying $n$ and $J$, we map out the ground state phase diagram. Figure~\ref{fig:Jdiag} shows the variational phase diagrams calculated for the effective spin model and the DE models. Figure~\ref{fig:Jdiag}(a) shows the result for the effective spin model constructed from the strong coupling theory [Eq.~(\ref{eq:Heff})]. For all $1/4<n<1/2$, we found both SC and 32-sub phases between the 2I2O and AIAO phases.

On the other hand, the DE model shows a slightly different phase diagram. Figure~\ref{fig:Jdiag}(b) shows the phase diagram for the model in Eq.~(\ref{eq:HDE2}). In Fig.~\ref{fig:Jdiag}(b), we found only the 32-sub state in the intermediate range of $J$ for $0.321\lesssim n\lesssim 0.466$, while only the SC state appears for $n\lesssim0.288$ and $n\gtrsim0.471$; the two phases appear in the narrow range of $0.288\lesssim n\lesssim 0.321$ successively while increasing $J$. Between these phases as well as to the 2I2O and AIAO phases, the system exhibits an electronic phase separation, which is due to the discontinuity in $n$ associated with the first-order magnetic phase transition. The result indicates that, with varying $n$, there are two regions in the phase diagram where the phase competing region is dominated by either 32-sub or SC order. This result is in contrast to the strong coupling theory, where the two phases appear for all $n$. Nevertheless, the strong coupling theory predicts the trend of instability toward the correct intermediate ground states found in the DE model.

In the last, we discuss the variational phase diagram for the DE model in Eq.~(\ref{eq:HDE}), without approximating $t_{ij}$ by $\tilde{t}_{ij}$ in Eq.~(\ref{eq:tilde_tij}). The result is shown in Fig.~\ref{fig:Jdiag}(c). The SC phase appears only for $n\lesssim0.25$, while the 32-sub phase appears for $0.323\lesssim n\lesssim0.418$; no intermediate phase is found for $0.25\lesssim n \lesssim 0.323$ and $n\gtrsim0.418$. Hence, despite the absence of quantum phase in the hopping term, the two DE models show qualitatively similar ground state phase diagrams. This is a crucial observation for justifying the approximation we used in this paper, as we ignored the effect of the quantum phase in the strong coupling theory.

\subsection{Monte Carlo simulation} \label{ssec:result:mc}

For further comparison, we study the models in Eq.~(\ref{eq:HDE}) using the unbiased MC method, which allows calculating thermodynamic quantities of the model in Eq.~(\ref{eq:HDE}) without any approximations.~\cite{Yunoki1998} This method and its variants~\cite{Motome1999,Furukawa2004} have recently been used to explore unconventional phases in the phase competing region of the DE models on frustrated lattices.~\cite{Kumar2010,Venderbos2012,Ishizuka2013,Reja2015}

The calculations were done with the system size $N=4\times N_{\rm s}$ with $N_{\rm s}=4^3$ under the periodic boundary conditions. Thermal averages of physical quantities were calculated for typically 3600 MC steps after 600 steps for thermalization. Some of the low-temperature data were calculated for longer MC steps up to 10400 steps. We divided the MC measurements into five bins and estimated the statistical errors by the standard deviations among the bins.

\begin{figure}
   \includegraphics[width=.8\linewidth]{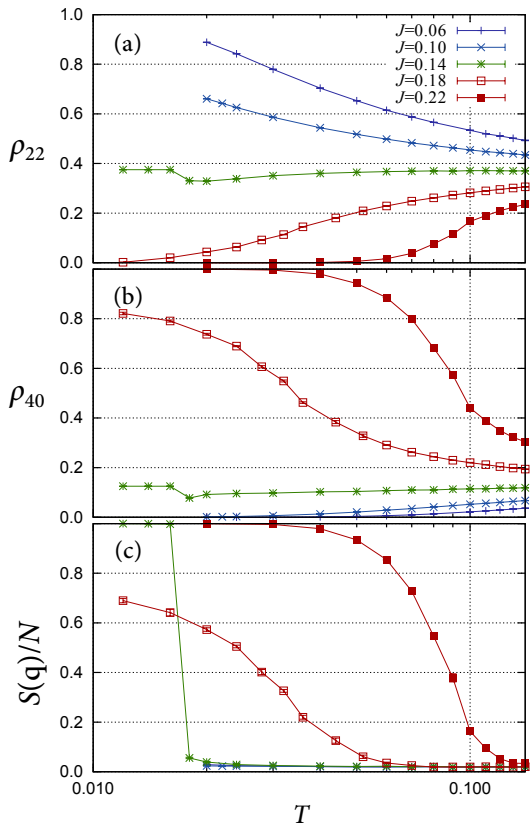}
   \caption{(Color online).
   Results of Monte Carlo simulation for the DE model in Eq.~(\ref{eq:HDE}) at $n\sim0.38$: (a) $\rho_{22}$, (b) $\rho_{40}$, and (c) $S({\bm q})/N$. See the text for details.
   }
   \label{fig:mc}
\end{figure}

MC results for both DE and effective spin models at $n=0.25$ have already been published by the authors,~\cite{Ishizuka2013} which are in accordance with the variational phase diagram in Fig.~\ref{fig:Jdiag}(c). In the DE model, only the SC state was found at the lowest temperature while increasing $J$, as the ground state in intermediate region. In addition, a fluctuating SC state with spatial inversion symmetry breaking was obtained in the intermediate temperature range above the critical temperature of the SC order. However, we did not find the 32-sub order in accordance with the variational phase diagram.

To check the existence of the 32-sub order in the region $n>1/4$, we here performed the MC simulation for the DE model in Eq.~(\ref{eq:HDE}). Figure~\ref{fig:mc} shows the result of MC simulation for $N=4\times 4^3$ sites at $n=0.38$. Figures~\ref{fig:mc}(a) and \ref{fig:mc}(b) show ratio of tetrahedra with two-in two-out ($\rho_{22}$) and all-in/all-out ($\rho_{40}$) configurations. When $J=0.06$, the results show enhancement of $\rho_{22}$ and decrease of $\rho_{40}$ with decreasing temperature. This is a sign of the FM correlation for the NN bonds.~\cite{dHertog2000,Bramwell2001} However, we could not reach the transition temperature for the long-range order in our simulation due to the freezing at lower temperatures. In contrast, for $J=0.22$, $\rho_{40}$ increases with decreasing temperature, approaching $\rho_{40}\to1$. In addition, the structure factor
\begin{eqnarray}
S_\alpha({\bm q}) = \frac1N\sum_{i,j\in\alpha} \langle {\bm S}_i\cdot{\bm S}_j\rangle \exp[{\rm i}{\bm q}\cdot({\bm r}_i-{\bm r}_j)]
\end{eqnarray}
for ${\bm q}=\bm{0}$ shows increase from zero as shown in Fig.~\ref{fig:mc}(c), indicating the AIAO order in the ground state.

On the other hand, the result for $J=0.14$ shows $\rho_{22}\to0.375$ and $\rho_{40}\to0.125$ with decreasing temperature [Figs.~\ref{fig:mc}(a) and \ref{fig:mc}(b)]. This is a sign of the phase transition to 32-sub order, where $6/16$ ($2/16$) tetrahedra are in 2I2O (AIAO) spin configuration.~\cite{Ishizuka2012} This transition is also confirmed by the spin structure factor for ${\bm q}=(\pi,\pi,\pi)$ plotted in Fig.~\ref{fig:mc}(c).

From these results, we conclude that, a sequence of states, 2I2O, 32-sub, and AIAO, appears for $n=0.38$ with increasing $J$. On the other hand, at $n=0.25$, the presence of SC order in the intermediate $J$ region was previously reported,~\cite{Ishizuka2013} consistently with the variational phase diagram in Fig.~\ref{fig:Jdiag}(c). The results support that the effective spin model by the strong coupling theory can predict the trend of intermediate phases in the competing region in the DE model.

\section{
Extension to Heisenberg and $XY$ Moments
}\label{sec:heisenberg}

In the last, we briefly discuss potential extension of this method to DE models with Heisenberg or $XY$ type localized moments. In these cases, we cannot use the expansion in Eq.~(\ref{eq:tilde_tij}). One alternative approach is to replace $\cos\theta_{ij}$ in Eq.~(\ref{eq:transfer}) by ${\bm S}_i\cdot{\bm S}_j$, and consider $\tilde{t}_{ij}-\alpha t_{ij}^\text{(bare)}$ as the perturbation, where $0<\alpha<1$. In the leading order, this gives FM NN interaction of the form $|J_{ij}|\propto\tilde{t}_{ij}$; the classical spin model with $J_{ij}$ as the NN interaction was studied motivated by the DE models.~\cite{Caparica2000}

Another route to evaluate the effective interactions is to use an asymptotic expansion, for instance, Taylor series
\begin{eqnarray}
\tilde{t}_{ij}
&=&\frac{t}{\sqrt{2}}\sqrt{1+{\bm S}_i\cdot{\bm S}_j},\\
&=&\frac{t}{\sqrt{2}}\left\{1+\frac{{\bm S}_i\cdot{\bm S}_j}2-\frac{({\bm S}_i\cdot{\bm S}_j)^2}8+\cdots\right\}.
\end{eqnarray}
This expansion naturally predicts the presence of positive biquadratic interaction as the subleading interaction. Recently, the positive biquadratic interaction has been proposed~\cite{Akagi2012} as the possible origin of non-coplanar phases found in the weak coupling region of a triangular Kondo lattice model.~\cite{Martin2008,Akagi2010,Kato2010} The same non-coplanar state has also been found in the DE limit, but its mechanism have not been explained so far.~\cite{Kumar2010} A simple argument based on our theory suggests that the non-coplanar phase in the DE limit may also be stabilized by the biquadratic interaction that arises from the strong coupling expansion. This also implies that, in the DE models with Heisenberg moments, even $O(t_{ij}^1)$ in the expansion may give rise to unconventional magnetism.

\section{Discussions and Summary}\label{sec:summary}

To summarize, in this paper, we studied the fermion-mediated effective spin interaction in an Ising spin double-exchange model on a pyrochlore lattice. To evaluate the effective interactions from microscopic theory, we used a strong coupling expansion and calculated the effective interactions up to second order in terms of the spin-dependent electron hopping. We showed that, effective four-spin interactions appear in the second order expansion. We also found that the effective interactions are limited to short range for this model, at least, for electron density $1/4\le n \le 1/2$.

Focusing on this region, we studied the accuracy of strong coupling theory. From comparison to the numerical results on the double-exchange model, we found the strong coupling theory gives a good estimate of the ground state energy. In addition, we studied the ground state phase diagram in the presence of antiferromagnetic super-exchange interactions between the localized moments. The calculations were done by a variational method, and some of the results were also confirmed by a Monte Carlo simulation. We found that the strong coupling method correctly captures the trend of intermediate phases found in the double-exchange models.

It is interesting that the expansion up to second order appears to capture the correct trend of the DE model, although the energy scale we are discussing is very small, $E/t\sim0.01$. This may be related to the fact that the ground state of a spin model is often sensitive only to the sign and not to the magnitude of interactions. For the effective spin model we studied, the 32-sublattice and spin-cluster states appear in the region of both second- and third-neighbor exchange interactions being antiferromagnetic.~\cite{Ishizuka2013} Hence, although the second order in expansion is not sufficient in correctly predicting which phase wins, it is still successful in guessing the correct candidates for the intermediate phase. This arguments cast a possible restriction on the application of the strong coupling theory we proposed; the strong coupling theory predicts correct trend only when the double-exchange model has a few subdominant interactions that are relevant, in addition to the ferromagnetic nearest-neighbor interaction. However, if this condition is satisfied, we can expect that the theory gives the correct trend. 

\acknowledgements

The authors thank N. Furukawa and M. Udagawa for fruitful discussions. This research was supported by KAKENHI (No. 24340076), the Strategic Programs for Innovative Research (SPIRE), MEXT, and the Computational Materials Science Initiative (CMSI), Japan. HI is supported by JSPS Postdoctoral Fellowships for Research Abroad.


\begin{references}
\bibitem{note_nophase}     The strong coupling expansion used here can be worked out without using the approximation in Eq.~(\ref{eq:transfer2}). In the present study, however, we use this approximation for simplicity. In Sec.~\ref{sec:result}, we will compare the result of strong coupling theory to the numerical results both with and without the approximation.
\bibitem{note_variational} We also evaluated the ground state energy for ${\bm q}=(0,0,2\pi)$ stacking of the 2I2O tetrahedra. This state may have a lower ground state energy than the $q=0$ 2I2O, but the relative difference is in the order of $\Delta E/ E=0.001$. Hence, it should not affect the presence of intermediate phases in a qualitative way.
\end{references}
\end{document}